\documentclass[10pt]{article}

\usepackage{amssymb,amsmath}
\usepackage{graphics}

\usepackage{amsmath}
\usepackage{amssymb}
\usepackage{txfonts}
\usepackage{mathrsfs}
\usepackage{cite}
\usepackage{mathtext}
\usepackage{epstopdf}
\usepackage{epsfig}
\usepackage{color}
\usepackage{amsfonts}

\usepackage{psfrag}

\usepackage{amscd}

\usepackage{epsfig,amsfonts,amssymb}
\usepackage{amsmath}
\usepackage{amssymb}
\usepackage{txfonts}
\usepackage{mathrsfs}
\usepackage{cite}
\usepackage{mathtext}
\usepackage[english]{babel}
\usepackage[latin1]{inputenc}
\usepackage{times}
\usepackage[T1]{fontenc}
\begin{document}

\begin{center}
{\Large A Covariant Form of the Navier-Stokes Equation for the Infinite Dimensional Galilean Conformal Algebra}
\end{center}
\centerline{\large \rm Ayan Mukhopadhyay}

\centerline{\large \it Harish-Chandra Research Institute}

\centerline{\large \it  Chhatnag Road, Jhusi,
Allahabad 211019, INDIA}
\vspace*{1.0ex}

\centerline{E-mail: ayan@mri.ernet.in}

\begin{abstract}
We demonstrate that the Navier-Stokes equation can be covariantized under the full infinite dimensional Galilean Conformal Algebra (GCA), such that it reduces to the usual Navier-Stokes equation in an inertial frame. The covariantization is possible only for incompressible flows, i.e when the divergence of the velocity field vanishes. Using the continuity equation, we can fix the transformation of pressure and density under GCA uniquely. We also find that when all chemical potentials vanish, $c_{s}$, which denotes the speed of sound in an inertial frame comoving with the flow, must either be a fundamental constant or given in terms of microscopic parameters. We will discuss how both could be possible. In absence of chemical potentials, we also find that the covariance under GCA implies that either the viscosity should vanish or the microscopic theory should have a length scale or a time scale or both.  We further find that the higher derivative corrections to the Navier-Stokes equation, can be covariantized, only if they are restricted to certain possible combinations in the inertial frame. We explicitly evaluate all possible three derivative corrections. Finally, we argue that our analysis hints that the parent relativistic theory with relativistic conformal symmetry needs to be deformed before the contraction is taken to produce a sensible GCA invariant dynamical limit.

\end{abstract}

\newpage \baselineskip=18pt \setcounter{footnote}{0}
\tableofcontents
\section{Introduction}
A new non-relativistic extension of the AdS/CFT conjecture \cite{adscft} became possible when it was shown \cite{GCA, 0902.1385} that a non-relativistic conformal algebra could be obtained as a parametric contraction of the relativistic conformal group. This contraction retained the same number of generators as the relativistic conformal group. It was also found out by the authors of \cite{0902.1385} that an inifinite-dimensional extension of the finite non-relativistic algebra was possible and following them, we call this algebra the Galilean Conformal Algebra, in short GCA. In the context of developing theversion of AdS/CFT for this non-relativistic symmetry, important steps were also taken in \cite{0902.1385} and later these have been extended in \cite{Alishahiha,0903.5184} (for some related work, please also see \cite{0904.0531} \footnote{Superconformal extensions have been dealt with in \cite{superconf}.}). The development is still under progress, however it has been realized that this is different from the case of the non-relativistic Schrodinger group. The Schrodinger group, has the advantage that, it can be embedded in the relativistic conformal group of two higher dimensions, so AdS/CFT in this case, can be developed on lines closer to the conventional relativistic setting, though in two higher dimensions \cite{Schrodinger}. In the case of the Galilean Conformal Algebra, however, it seems that the dynamics in the bulk involves a degenerate limit, which is possibly a Newton-Cartan like gravity involving an $AdS_{2}$ factor \cite{0902.1385} \footnote{For some interesting earlier work, please look at \cite{Plyuschay}.}.

To get a better understanding, it will be useful to understand the pure gravity sector first and in this sector, the gravity duals of hydrodynamic flows ubiquitously plays a very special role, because of the conceptual clarity of their construction (for a review, see \cite{Rangamani}). However, even before constructing gravity duals, it is important, to understand the role of the \textit{full} Galilean Conformal Algebra as symmetries of the hydrodynamics of the boundary theory. In the originl work \cite{0902.1385}, it was shown that the Euler equation for incompressible flows was invariant under some of the elements of the Galilean Conformal Algebra. However, the hydrodynamics in any physical theory, should have a non-zero viscosity \footnote{In fact there is a conjectured lower bound on the viscosity originally due to KSS \cite{KSS}. For a recent review, please see \cite{Teaney}.} and moreover there are typically higher derivative corrections to all orders. Here, we will investigate how the Galilean Conformal Algebra can act as symmetries of the Navier-Stokes equation and also its role in constraining higher derivative corrections.   

The important point of our approach will be that we will be looking for \textit{covariance} rather than \textit{invariance}, in close analogy with the case of relativistic conformal hydrodynamics where the relativistic Navier-Stokes equation and its higher derivative corrections can be made covariant (not invariant) under the relativistic conformal group \cite{Shiraz}. An element in GCA may take a Galilean inertial frame to a non-inertial one. After \textit{covariantizing} under GCA, as expected, the equation will take its usual form in an inertial frame, but in a non-inertial frame it will assume a non-standard form. In our case, the covariantizing will involve novel features, like the absolute (time-dependent) acceleration and absolute (time-dependent) angular velocity of the non-inertial frame,\textit{which are not non-relativistic degenerations of the relativistic covariant form.}. The basic reason for the appearance of novel features is straightforward, the infinite GCA has no relativistic analogue (for a lucid description of non-relativistic degenerations of relativistically covariant hydrodynamics, etc, please see \cite{Kunzle}). Also, in non-relativistic dynamics, the absolute acceleration or the absolute angular velocity of a non-inertial frame are the more natural objects to be used for  covariantizing rather than "connections". Since our approach involves covariantizing the usual Navier-Stokes equation for incompressible flows which holds in inertial frames, it is very different from that in \cite{main}\footnote{For some related work please also see \cite{sub}.}. 

We will divide the Navier-Stokes equation into three parts, namely, the kinematic term, the pressure term and the viscous term, and we will show that each term separately transforms covariantly, exactly like in the case of the covariance of the relativistic Navier-Stokes equation under the relativistic conformal group. The kinematic term, in an inertial frame, is just the Euler derivative acting on the velocity field. This term transforms just like the acceleration. Since, the GCA can transform an inertial frame to a non-inertial frame, as mentioned above, the covariantizing will naturally involve the absolute angular velocity and the absolute acceleration of the non-inertial frame. However, the covariance under the ``spatially correlated time reparametrizations'' will be possible only if the flow is incompressible\footnote{When a non-relativistic limit is taken by applying an appropriate scaling of the relativistic Navier-Stokes equation, the incompressibility of the flow is automatically obtained (please see the first two references of \cite{main}. The GCA covariant form, however, cannot be obtained as a limit of the usual conformally covariant relativistic Navier-Stokes equation.}. Therefore, we would require the flow to be incompressible too.

The pressure term is just the gradient of the pressure divided by the density. We will show that this leads to the speed of sound being GCA invariant, essentially because the pressure transforms in the same way as the density under GCA. 

The viscous term is $(1/\rho)\partial_i \Pi_{ij}$, where $\rho$ is the density and $\Pi_{ij}$ is the shear stress tensor given by, $\Pi_{ij} = \eta (\nabla_i v_j + \nabla_j v_i -(2/3) \delta_{ij} \nabla\cdot v)$, with $\eta$ being the shear viscosity. Here the shear viscosity also transforms as a \textit{field} only through its dependence on the thermodynamic variables which transform under GCA. 

We will see that when all chemical potentials vanish (as in a gas of phonons), $c_{s}$, which denotes the speed of sound in an inertial comoving with the flow, is invariant under GCA. We will see that this implies that it must be a fundamental constant like the speed of light or given in terms of the microscopic parameters. We will see how each could be possible, in particular we will see that when the number of spatial dimensions is two, GCA admits a central charge with dimension $(1/speed)^2$. Then we will study the transformation of viscosity under GCA and see that in the absence of chemical potentials the transformation could be realized only if the microscopic theory contains a length scale, or a time scale, or both and if this is not possible, the viscosity should vanish.

We also find that the GCA also has the potential to restrict the possible corrections to the Navier-Stokes equation and we explicitly evaluate the possible three derivative corrections. It is intriguing that all these four possibilities correspond to the relativistic conformal case so that the relativistic terms reduce to our terms in the non-relativistic limit in inertial frames, when the flow is incompressible. The general lesson is that a phenomenological law can be covariantized under GCA only if its form in the inertial frame is sufficiently restricted. 

The plan of the paper is as follows. In section 2, we arrive at a covariant description of the hydrodynamics for the GCA. In section 3, we use this to covariantize the Navier-Stokes equation. In scetion 4, we discuss how we can covariantize the continuity equation and how it influences the transformations of the density, pressure and viscosity. In section 5, we show how the GCA constrains higher derivative corrections to the Navier-Stokes equations. Then we conclude with some discussions on the implications of our results for the version of AdS/CFT with GCA as the conformal symmetry group. In the appendices, we elucidate some technical points and in particular, we also give a simple mathematical interpretation of the GCA, that could be useful for constructing GCA invariant microscopic theories.

\section{Covariant Kinematics for the Infinite Galilean Conformal Algebra}
The finite part of the Galilean Conformal Algebra can be obtained as a parametric contraction of the $SO(d+1,2)$ relativistic conformal group of $(d,1)$ dimensional Minkowskian space-time \cite{GCA, 0902.1385}. This finite part forms a Lie group with exactly the same number of generators as the $SO(d+1,2)$ relativistic conformal group. The generators of this finite part consists of the following
\begin{eqnarray}
H = -\frac{\partial}{\partial t},\\\nonumber
P_{i} = \nabla_{i,}\\\nonumber
J_{ij} = -(x_{i}\nabla_{j}-x_{j}\nabla_{i}),\\\nonumber
B_{i} = t\nabla_{i},\\\nonumber
D = -(\mathbf{x.\nabla}+t\frac{\partial}{\partial t}),\\\nonumber
K = -(2t\mathbf{x.\nabla}+t^{2}\frac{\partial}{\partial t}),\\\nonumber
K_{i} = t^{2}\nabla_{i}.\\\nonumber
\end{eqnarray}
Clearly, $H$ is the Hamiltonian, $P_{i}$ are the momentae and $J_{ij}$ are the angular momentae generating time translations, spatial translations and angular rotations respectively. The $B_{i}$'s generate the Galilean boosts. The dilation operator $D$ acts differently from the Schrodinger group as it scales all spatial coordinates and time in the same way. The other generators $K$ and $K_{i}$ can be thought of non-relativistic counterparts of relativistic special conformal transformations.
 
This finite algebra has an infinite extension which forms the full GCA, the generators of which can be labelled as below
\begin{eqnarray}
L^{(n)}= -(n+1)t^{n}(\mathbf{x.\nabla})-t^{n+1}\frac{\partial}{\partial t},\\\nonumber
M_{i}^{(n)} = t^{n+1}\nabla_{i},\\\nonumber
J_{a}^{(n)} \equiv J_{ij}^{(n)} = -t^{n}(x_{i}\nabla_{j}-x_{j}\nabla_{i}),
\end{eqnarray}
where n runs over all integers. The $SL(2,R)$ part of $L^{(n)}$'s belong to the finite group (as $H=L^{(-1)}, D= L^{(0)}, L^{(1)}=K$). Also, $P_{i} = M_{i}^{(-1)}, B_{i} = M_{i}^{(0)}, K_{i} = M_{i}^{1}$, while only $J_{ij}^{(0)}$ belong to the finite group. The full algebra is
\begin{eqnarray}
[L^{(m)},L^{(n)}] = (m-n) L^{(m+n)},\\\nonumber
[L^{(m)},J_{a}^{(n)}]=-nJ_{a}^{(m+n)},\\\nonumber
[J_{a}^{(n)},J_{b}^{(m)}] = f_{abc}J_{c}^{(n+m)},\\\nonumber
[L^{(m)}, M_{i}^{(n)}]=(m-n)M_{i}^{(m+n)},\\\nonumber
[J_{ij}^{(n)},M_{k}^{(m)}] = -(M_{i}^{(m+n)}\delta_{jk}-M_{i}^{(m+n)}\delta_{jk}),\\\nonumber
[M_{i}^{(m)},M_{j}^{(n)}] = 0.
\end{eqnarray}
The index $a$ above form an alternative label corrsponding to the spatial rotation group $SO(d)$ and $f_{abc}$ are the structure constants of this group. Further $J_{(a)}^{(n)}$'s and $L^{(m)}$'s together form a Virasoro Kac-Moody algebra. The GCA admits the usual (dimensionless) central charges for the Virasoro Kac-Moody subalgebra as the $M_{i}^{(n)}$'s can be consistently put to zero \cite{0902.1385}. Besides, these usual dimensionless, central charges, a special kind of central charge, is possible in the case of two spatial dimensions and it will be important for us because only in the case of two spatial dimensions we can have a dimensionful central charge. A dimensionful central charge, unlike a dimensionless one can appear in the Lagrangian description of the theory. A simple example is the central charge with dimension of mass in the Schrodinger group actually being the mass of the free particle. This central charge $\Theta$, appears in the commutator of $M_{i}^{(m)}$'s in the GCA as below \cite{0903.5184,Lukierski1}
\begin{equation}
[M_{i}^{(m)},M_{j}^{(n)}]=I^{mn}\epsilon_{ij}\Theta,
\end{equation}
where $I^{mn}$ is the invariant tensor of the spin one representation of $SL(2,R)$. The central charge $\Theta$ has the dimension of $(1/speed)^{2}$. For possible physical interpretations of this term, please look at \cite{0903.5184,Lukierski1,2+1D}. Further, in the case of the Schrodinger group, as mentioned above, there is another possible central charge (for any number of spatial dimensions) which has the dimension of mass (in units where the Planck's constant is set to unity, mass is basically time divided by square of length) and in fact has the interpretation of the mass scale in the corresponding theory. The absence of this central term in the GCA has been argued \cite{0902.1385,0904.0531} to reflect the absence of any mass scale in the microscopic theory and we will also hold to this point of view here. 

The $J_{a}^{(n)}$'s actually generate arbitrary time dependent rotations, the $M_{i}^{(n)}$'s generate arbitrary time-dependent boosts and the $L^{(n)}$'s generate spatially correlated time reparametrisations \cite{0902.1385}. Each of these form a subalgebra by themselves. We now proceed to consider each of these categories of space-time transformations in detail to see how one can have a covariant description of kinematics for each of these categories. Finally, we will sum up by arriving at a kinematic description which will be covariant under the full set of transformations.
\subsection{Arbitrary time dependent rotations}
These transformations are
\begin{eqnarray}\label{rot}
x_{i}^{'} = R_{ij}(t)x_j,\\\nonumber
t^{'} = t,
\end{eqnarray}
where $R_{ij}$ is an arbitrary time dependent rotation matrix (so that $R_{ij}^{-1}= R_{ji}$). The velocity transforms in the following manner,
\begin{equation}\label{velrot}
v_{i} = R_{ij}^{-1}(v_{j}^{'} - \frac{dR_{jk}}{dt^{'}}R_{kl}^{-1}x_{l}^{'}).
\end{equation}
Now we will show that from the above transformation one can extract a covariant time derivative. Let us define $\Omega_{ij}$ to be the absolute angular velocity of the non-inertial frame with respect to any inertial frame (note when the number of spatial dimensions is more than three this is actually a tensor, but by abuse of notation we will still call it absolute angular velocity, in three dimensions $\Omega_{ij} =\epsilon_{ikj}\Omega_{k}$). Suppose the unprimed coordinates are in the inertial frame and the primed ones are in the non-inertial frame. Then clearly the absolute angular velocity $\Omega_{ij} = -(dR_{ik}/dt)R_{kj}^{-1}$. Of course the absolute angular velocity of a frame is very much a physical quantity as it can be determined by an observer using that frame. The covariant time derivative in a given frame, can now be defined through its action on vectors as below,
\begin{equation}\label{covd}
\frac{D}{Dt}V_{i} = \frac{d}{dt}V_{i} + \Omega_{ij}V_{j},
\end{equation}
where $\mathbf{V}$ is an arbitrary vector. Note that in an inertial frame $D/Dt = d/dt$, so if the unprimed coordinates are inertial and primed coordinates non-inertial we may rewrite (\ref{velrot}) as,
\begin{equation}
\frac{D}{Dt}x_{i} = R_{ij}^{-1}\frac{D}{Dt^{'}}x_{j}^{'}.
\end{equation}
In fact, we may replace the position vector $x_i$ above with any arbitrary vector $V_i$ which transforms like $V_{i}^{'} = R_{ij}V_{j}$, then it also follows that
\begin{equation}\label{Vrot}
\frac{D}{Dt}V_{i} = R_{ij}^{-1}\frac{D}{Dt^{'}}V_{j}^{'}.
\end{equation}
We now claim that the above relation is valid even when both the primed and unprimed coordinates are non-inertial. An easy way to prove this is as follows. Let us take two non-inertial frames ($\mathbf{x_{(1)}}, t_{(1)}$) and ($\mathbf{x_{(2)}}, t_{(2)}$) which are related to the inertial frame ($\mathbf{x}, t$) through $x_{(1)i}=R_{(1)ij}x_{j}, t_{(1)} = t$ and $x_{(2)i}=R_{(2)ij}x_{j}, t_{(2)} = t$ respectively. Obviously the absolute angular velocities of the non-inertial frames are $\Omega_{(1)ij} = -(dR_{(1)ik}/dt)R_{(1)kj}^{-1}$ and $\Omega_{(2)ij} = -(dR_{(2)ik}/dt)R_{(2)kj}^{-1}$ respectively. Clearly,
\begin{equation}
\frac{D}{Dt}V_{i} = R_{(1)ij}^{-1}\frac{D}{Dt_{1}}V_{(1)j} = R_{(2)ij}^{-1}\frac{D}{Dt_{2}}V_{(2)j}.
\end{equation}
Therefore,
\begin{equation}
\frac{D}{Dt_{1}}V_{(1)i} = R_{ij}^{-1}\frac{D}{Dt_{2}}V_{(2)j},
\end{equation}
where
\begin{equation}
R_{ij} = R_{(2)ik}R_{(1)kj}^{-1},
\end{equation}
as required so that indeed $x_{(2)i}=R_{ij}x_{(1)j}$. Therefore, (\ref{Vrot}) is valid for any two frames, even if both are non-inertial. In particular we will define the covariant velocity $\mathbf{\mathcal{V}}^{(rot)}$ as the covariant derivative of the position vector so that
\begin{equation}
\mathcal{V}_{i}^{(rot)} = \frac{D}{Dt}x_{i} = \frac{d}{dt} x_{i} + \Omega_{ij}x_{j} .
\end{equation}
By construction this transforms covariantly under (\ref{rot}), so that
\begin{equation}
\mathcal{V}_{i}^{(rot)} = R_{ij}^{-1}\mathcal{V}_{j}^{(rot)'}.
\end{equation} 
The above tells us how to modify the acceleration so that we get a covariant vector. We define, $\mathbf{\mathcal{A}}^{(rot)}$ the ``covariant accelaration'' as two covariant time derivatives acting on the position vector as below,
\begin{equation}\label{accltn}
\mathcal{A}_{i}^{(rot)} = \frac{D^{2}}{Dt^{2}}x_{i} = \frac{d^{2}}{dt^{2}}x_{i} + 2\Omega_{ij}v_{j} + \Omega_{ij}\Omega_{jk}x_{k} + (\frac{d}{dt}\Omega_{ij})x_{j}.
\end{equation}
In the non-inertial coordinates in the right hand side of the last expression above the corrections to the usual acceleration are just the Corriolis, centrifugal and Euler forces respectively. \footnote{Usually the relation between acceleration in inertial frame and non-inertial frames in the case of three spatial dimensions are written from the ``passive'' point of view as: $\mathbf{a}^{'} = \mathbf{a} - 2 \mathbf{\Omega} \times \mathbf{v} + \mathbf{\Omega} \times (\mathbf{\Omega} \times \mathbf{x}) - (d\mathbf{\Omega}/dt)\times\mathbf{x}$, where the primed coordinates are non-inertial and unprimed ones are inertial. However, one can work out that it is, in fact, equivalent to $\mathbf{a} = \mathbf{a}^{'} + 2 \mathbf{\Omega} \times \mathbf{v}^{'} + \mathbf{\Omega} \times (\mathbf{\Omega} \times \mathbf{x}^{'}) + (d\mathbf{\Omega}/dt^{'})\times\mathbf{x}^{'}$. In three spatial dimensions this is just another way of understanding (\ref{accltn}).} By construction, under the transformations (\ref{rot}), the covariant acceleration transforms as below,
\begin{equation}
\mathcal{A}_{i}^{(rot)} = R_{ij}^{-1}\mathcal{A}^{(rot)'}_{i},
\end{equation}
where both the primed and unprimed coordinates can be non-inertial. 

We also observe that the spatial derivative $\nabla_i$ and the symmetric traceless tensor $\sigma_{ij} = \nabla_i v_j + \nabla_j v_i -(2/3) \delta_{ij} (\mathbf{\nabla.v})$ transforms covariantly while divergence of the velocity $\mathbf{\nabla\cdot v}$ transforms invariantly (in the last two cases, of course, we are talking of a velocity field), so that under the transformations (\ref{rot}),
\begin{eqnarray}
\nabla_{i} = R_{ij}^{-1} \nabla_{j}^{'},\\\nonumber
\sigma_{ij} = R_{ik}^{-1}R_{jl}^{-1}\sigma_{kl}^{'},\\\nonumber
\mathbf{\nabla\cdot v} = \mathbf{\nabla}^{'}\cdot\mathbf{v}^{'} .
\end{eqnarray}
For the last two results above, we have used the fact that $(dR_{ik}/dt)R_{kj}^{-1}$ is antisymmetric in $i$ and $j$. 

To summarize we see that we have two basic operators which transform covariantly, namely the covariant time derivative $D/Dt$ (as defined in (\ref{covd})) and the spatial derivative $\nabla_{i}$. Further the traceless symmetric tensor $\epsilon_{ij}$ transforms covariantly and $\mathbf{\nabla\cdot v}$ transforms invariantly.

\subsection{Arbitrary time dependent boosts}
These transformations are
\begin{eqnarray}\label{boost}
x_{i}^{'} = x_{i} + b_{i}(t),\\\nonumber
t^{'} = t.
\end{eqnarray}
We will mathematically interpret the above as the position vector not transforming covariantly. It is easy to see that relative distances, relative velocities and relative accelerations will remain invariant under these transformations. So, one can easily get a invariant acceleration field using the relative acceleration with respect to the absolute acceleration of the frame. Let $\mathbf{\mathcal{B}}$ be the absolute acceleration of the non-inertial frame. Then the invariant acceleration field $\mathbf{\mathcal{A}}^{(accl)}$ may be defined as,
\begin{equation}\label{accboost}
\mathcal{A}^{(accl)}_{i} = \frac{d}{dt}v_{i} - \mathbf{\mathcal{A}}^{(accl)}=  \frac{d}{dt}v_{i} - \nabla_{i}(\mathbf{\mathcal{B}.x}).
\end{equation}
This, again, can be proved as before, consider the unprimed coordinates as inertial and primed coordinates as non-inertial in (\ref{boost}), then from the passive point of view, the absolute acceleration of the non-inertial frame is $\mathcal{B}_{i} = d^{2} b_{i}/dt^{2}$. So it is clearly true that
\begin{equation}
\mathcal{A}^{(accl)}_{i} = \frac{d}{dt}v_{i}=\frac{d}{dt^{'}}v_{i}^{'} - \nabla_{i}^{'}(\mathbf{\mathcal{B}\cdot x^{'}})= \mathcal{A}^{(accl)'}_{i}.
\end{equation}
We can repeat the same trick of comparing two non-inertial frames with one inertial frame and then comparing the two non-inertial frames with each other, as desribed in the previous subsection, to conclude that $\mathcal{A}^{(accl)}_{i} = \mathcal{A}^{(accl)'}_{i}$ is valid even if both the primed and unprimed frames are non-inertial. Therefore we conclude that (\ref{accboost}) indeed defines an invariant acceleration field.

We also observe the operator $\nabla_{i}$ is invariant and so are $\mathbf{\nabla\cdot v}$ and the symmetric traceless tensor $\epsilon_{ij}$ under the transformation (\ref{boost}). 

\subsection{Spatially correlated time reparametrizations}
These transformations are
\begin{eqnarray}\label{sctr}
x_{i}^{'} = \frac{df}{dt}x_{i},\\\nonumber
t^{'}= f(t).
\end{eqnarray}
The interesting thing about this transformation is that the new frame may be using a different time from absolute time. However, one must ask how can an observer using a frame know that the time being used is different from absolute time? To find that out, let us first note the transformation of the velocity, 
\begin{equation}\label{velsc}
v_{i} =v_{i}^{'} + \frac{\frac{d^2 t}{dt^{'2}}}{\frac{dt}{dt^{'}}} x_{i}^{'}.
\end{equation}
The divergence of the velocity field transforms as,
\begin{equation}
\mathbf{\nabla\cdot v}= \frac{dt^{'}}{dt}\mathbf{\nabla^{'}\cdot v^{'}} + d \frac{\frac{d^2 t}{dt^{'2}}}{(\frac{dt}{dt^{'}})^2}.
\end{equation}
Combining these one can easily see that one can make an invariant velocity field, 
\begin{equation}\label{velin}
\mathcal{V}^{(sctr)}_{i} = v_{i} - \frac{\mathbf{\nabla\cdot v}}{d}x_{i}.
\end{equation}
Firstly let us assume that when the frame is using absolute time the divergence of the velocity field, $\mathbf{\nabla\cdot v}$ vanishes. After a generic transformation as in (\ref{sctr}), as shown in (\ref{velsc}), clearly it will no longer be zero. Therefore, if is this is not zero, one knows that the time being used is not using absolute time. Note the divergence of the velocity field remains zero under constant dilatation or shifts, so one can be sure of the use of absolute time only upto a constant dilation or shift. Now, (\ref{velin}) shows that \textit{one can construct an invariant velocity field under space correlated time reparametrisations, which reduces to the usual velocity field in an inertial frame (where absolute time is used), if and only if, the divergence of the velocity field vanishes (or the flow is incompressible) in the inertial frame.} This is precisely why the assumption of incompressible flow is crucial to covariantize the Navier-Stokes' equation under the full GCA.

One can make a covariant acceleration field
\begin{equation}
\mathcal{A}^{(sctr)}_{i} = \frac{d}{dt}\mathcal{V}^{(sctr)}_i,
\end{equation}
so that
\begin{equation}
\mathcal{A}^{(sctr)}_{i} = \frac{dt^{'}}{dt}\mathcal{A}^{(sctr)'}_{i}.
\end{equation}

Finally one notes that the operators $\nabla_i$ transforms covariantly and so does the traceless symmetric tensor $\sigma_{ij}$;
\begin{eqnarray}
\nabla_{i} = \frac{dt^{'}}{dt}\nabla^{'}_{i},\\\nonumber
\sigma_{ij} = \frac{dt^{'}}{dt}\sigma_{ij}.
\end{eqnarray}

\subsection{Summing all up}
We would like to sum up all our results in order to construct a covariant acceleration field which will be covariant under the full GCA. We first observe that any element of the GCA can be written as a succession of a time dependent rotation, a spatially correlated time reparametrisation and a time dependent boost (for proof please see appendix B). So without loss of generality, any element of GCA can be written as below:
\begin{eqnarray}\label{comb}
x_{i}^{'}= \frac{df}{dt}R_{ij}(t) x_{j}+ b_{i}(t),\\\nonumber
t^{'} = f(t).
\end{eqnarray}
Instead of working out what happens under the full transformation we can, instead, use the following logic. Let us first put $b_{i}(t)$ to zero so that the position vector transforms covariantly. Then one can define a velocity field which is covariant under the combined action of rotation and spatially correlated time reparametrization.
\begin{equation}\label{covvb=0}
\mathcal{V}^{(\mathbf{b}=0)}_i = v_i + \Omega_{ij}x_j - \frac{\mathbf{\nabla\cdot v}}{d}x_{i} .
\end{equation}
However, now the angular velocity of the frame $\Omega_{ij}$ is defined with the time of the frame, which need not be the absolute time, for instance, in (\ref{comb}), if the primed coordinates are non-inertial and the unprimed one is inertial then the angular velocity of the non-inertial frame is $\Omega_{ij} = -(dR_{ik}/dt^{'})R_{kj}^{-1}$. One can easily see the further modification which makes the velocity field covariant as $\mathbf{\nabla\cdot v}$ transforms invariantly under arbitrary rotations. Anyway, using methods pointed out in the previous subsections, one can readily check that when $b_{i}(t)=0$, under the transformation (\ref{comb}), the covariant velocity field transforms as,
\begin{equation}
\mathcal{V}^{(\mathbf{b}=0)}_{i} = R_{ij}^{-1}\mathcal{V}^{(\mathbf{b}=0)'}_{j}.
\end{equation}
If we have a vector $V_{i}$, which transforms under (\ref{comb}) when $b_{i}(t)=0$ as
\begin{equation}
V_{i} = R_{ij}^{-1}V_{j},
\end{equation}
then we define its covariant time derivative as
\begin{equation}\label{combd}
\frac{D}{Dt}V_{i} = \frac{d}{dt}V_{i} + \Omega_{ij}V_{j}.
\end{equation}
Then when $b_{i}(t)=0$, under the transformation (\ref{comb}) we get
\begin{equation}
\frac{D}{Dt}V_{i} = \frac{dt^{'}}{dt}R_{ij}^{-1}\frac{D}{Dt^{'}}V^{'}_{j}.
\end{equation}
The above can be easily proved by our previous trick of comapring two non-inertial frames with an inertial one and then comparing the non-inertial frames with each other so that the above remains valid even when both the primed and unprimed frames are non-inertial. For the sake of convenience of the reader, we will repeat this trick explicitly for our final covariant acceleration field, which we are now in the process of constructing. It is now clear how we should construct a covariant acceleration field when $b_{i}(t)=0$. We must make the covariant time derivative act on the covariant velocity field, so that,
\begin{equation}
\mathcal{A}^{(\mathbf{b}=0)}_{i}  = \frac{D}{Dt}\mathcal{V}^{(\mathbf{b}=0)}_{i} = \frac{d}{dt}(v_i + \Omega_{ij}x_j - \frac{\mathbf{\nabla.v}}{d}x_i) + \Omega_{ij}(v_j + \Omega_{jk}x_k - \frac{\mathbf{\nabla.v}}{d}x_j).
\end{equation}
Therefore, when $b_{i}(t)=0$, under the combined transformation (\ref{comb}), the covariant acceleration field constructed above transforms as $\mathbf{V}$ in (\ref{combd}), so that
\begin{equation}
\mathcal{A}^{(\mathbf{b}=0)}_{i} = \frac{dt^{'}}{dt}R_{ij}^{-1}\mathcal{A}^{(\mathbf{b}=0)'}_{j}.
\end{equation}
Again it is clear how we can maintain the above covariance when $b_{i}(t)$ is not zero. We just take the relative covariant acceleration with respect to $\mathbf{\mathcal{B}}$, the acceleration of the frame in the time of the frame (which may not be absolute time). Our final covariant acceleration field, which is covariant with respect to the full GCA is:
\begin{eqnarray}\label{acomb}
\mathcal{A}^{(comb)}_{i} = \mathcal{A}^{(\mathbf{b}=0)}_{i} - \mathcal{B}_{i} = \mathcal{A}^{(\mathbf{b}=0)}_{i} - \nabla_{i}(\mathbf{\mathcal{B}\cdot x}) = \frac{D}{Dt}\mathcal{V}^{(\mathbf{b}=0)}_{i} - \nabla_{i}(\mathbf{\mathcal{B}\cdot x})\\\nonumber
= \frac{d}{dt}(v_i + \Omega_{ij}x_j - \frac{\mathbf{\nabla.v}}{d}x_i) + \Omega_{ij}(v_j + \Omega_{jk}x_k - \frac{\mathbf{\nabla.v}}{d}x_j) - \nabla_{i}(\mathbf{\mathcal{B}\cdot x}).
\end{eqnarray}
The covariance, under the full GCA is simply
\begin{equation}\label{acov}
\mathcal{A}^{(comb)}_{i} = \frac{dt^{'}}{dt}R_{ij}^{-1}\mathcal{A}^{(comb)'}_{j}.
\end{equation}
To check the above, one can go back again to the representation (\ref{comb}) of an arbitrary element of GCA. Now let us suppose that the unprimed coordinates are inertial (where the time is absolute time) so that $\Omega_{ij}, \mathcal{B}_{i}, \mathbf{\nabla.v}$ are all zero in these coordinates. The covariant acceleration field is just the usual acceleration $dv/dt$ in these coordinates. Now one can readily check the validity of (\ref{acov}) with the definition (\ref{acomb}) of the covariant acceleration field with:
\begin{eqnarray}\label{relate}
\Omega_{ij} = -(\frac{d}{dt^{'}}R_{ik})R_{kj}^{-1}\\\nonumber
\mathcal{B}_{i} = \frac{D^{2}}{Dt^{'2}}b_{i}(t(t^{'})) = \frac{d^2}{dt^{'2}}b_{i} - 2 \Omega_{ij}\frac{d}{dt^{'}}b_{j} + \Omega_{ij}\Omega_{jk}b_{k} - (\frac{d}{dt^{'}}\Omega_{ij})b_{j}
\end{eqnarray}
The above relations are familiar in usual Galilean kinematics, except for the use of a general time $t^{'}$ in the non-inertial frame, which may not be the absolute time. Now as before we consider another non-inertial frame $(\mathbf{x^{''}}, t^{''})$ related to the same inertial frame $(\mathbf{x},t)$ through the same relation (\ref{comb}), but with different parameters ($R_{ij}^{'}(t), f^{'}(t), b_{i}^{'}(t))$. Then again (\ref{acov}) is valid with the deinition (\ref{acomb}) of the covariant acceleration field and with the angular velocities and acceleration of this frame given by (\ref{relate}), but $(R_{ij}, b_{i})$ replaced by $(R_{ij}^{'}, b_{i}^{'})$. As a result
\begin{equation}
\mathcal{A}^{(comb)}_{i} = \frac{dt^{'}}{dt}R_{ij}^{-1}\mathcal{A}^{(comb)'}_{j} =  \frac{dt^{''}}{dt}R_{ij}^{'-1}\mathcal{A}^{(comb)''}_{j}.
\end{equation}
The above implies
\begin{equation}
\mathcal{A}^{(comb)'}_{j}= \frac{dt^{''}}{dt^{'}}R_{ij}R_{jk}^{'-1}\mathcal{A}^{(comb)''}_{k} = \frac{dt^{''}}{dt^{'}}(R_{ij}^{'}R_{jk}^{-1})^{-1}\mathcal{A}^{(comb)''}_{k}.
\end{equation}
The last equality above is exactly what is required for the validity of (\ref{acov}) between these two non-inertial frames and since by choice they were arbitrary, we have proved that (\ref{acov}) is valid for any two coordinates. However, we note that the covariant acceleration field as defined in (\ref{acomb}) reduces to the usual acceleration field in an inertial frame only if the flow is incompressible in the inertial frame. So, we prove that \textit{it is possible to define a covariant acceleration field as defined in (\ref{acomb}) which transforms covariantly as in (\ref{acov}) under the full GCA if and only if the flow is incompressible (i.e. $\mathbf{\nabla\cdot v} =0$) in an inertial frame (where absolute time is used).}

Finally we note that the operator $\nabla_{i}$ transforms covariantly under the full GCA and so does the traceless symmetric tensor $\sigma_{ij}$. Under the transformation (\ref{comb})
\begin{eqnarray}\label{list}
\nabla_{i} = \frac{dt^{'}}{dt}R_{ij}^{-1}\nabla_{j}^{'},\\\nonumber
\sigma_{ij} = \frac{dt^{'}}{dt}R_{ik}^{-1}R_{jl}^{-1}\sigma_{kl}^{'}.
\end{eqnarray}
\section{Covariantizing the Navier-Stokes Equation}
The approach to equilibrium in physical systems is captured usually by three equation, namely, the continuity equation, the Navier-Stokes equation and the equation for evolution of the mean isotropic pressure. Of these three, the Navier-Stokes equation concerned with the approach to mechanical equilibrium is the most fundamental. The continuity equation is valid only if the microscopic interactions conserve particle number. When the flow is incompressible, i.e when the divergence of the velocity field (whose take values corresponding to the local mean particle velocity) vanishes, the pressure actually is not an independent dynamical variable as it does not have an independent equation for its evolution \cite{Landau}.  

As mentioned in the Introduction, we will dissect the Navier-Stokes equation into the kinematic term, the pressure term and the viscous term, and establish the covariance of each of these terms under GCA.

\subsection{The Kinematic Term}
The kinematic term, in an inertial frame, is simply $d\mathbf{v}/dt$, the acceleration field. Now the total time derivative $d/dt$ acting on any field is simply the Euler operator $\mathcal{D}=\partial/\partial_{t} + \mathbf{v.\nabla}$ acting on the field. Therefore the covariant form of the kinematic term, under the full GCA is just the covariant acceleration field (\ref{acomb}) where, we may replace $d/dt$ with $\mathcal{D}$
\begin{equation}\label{kin}
(\mathcal{D}v)^{(comb)}_{i}= \mathcal{D}(v_i + \Omega_{ij}(t)x_j - \frac{\mathbf{\nabla.v}}{d}x_i) + \Omega_{ij}(t)(v_j + \Omega_{jk}(t)x_k - \frac{\mathbf{\nabla.v}}{d}x_j) - \nabla_{i}(\mathbf{\mathcal{B}}(t)\cdot\mathbf{x}).
\end{equation}
Above we have made explicit that the angular velocity and acceleration of the frame is time dependent only. As we have proved in the previous section, the kinematic term transforms as (\ref{acov}) under the full GCA, so under the transformation (\ref{comb}), the covariant acceleration field transforms as
\begin{equation}
 (\mathcal{D}v)^{(comb)}_{i}= \frac{dt^{'}}{dt}R_{ij}^{-1}(\mathcal{D}v)^{(comb)'}_{i}.
\end{equation}
Note the covariant kinematic term (\ref{kin}) becomes the usual kinematic term in an inertial frame, where absolute time is also used, only when the flow is incompressible in any inertial frame. So, it is crucial that the flow, is indeed, incompressible, in an inertial frame. \textit{The kinematic term can be made GCA covariant only if the flow is incompressible in an inertial frame so that it reduces to just the Euler derivative acting on the velocity field in an inertial frame.} 

We also note that since the centrifugal force is a conservative force, one may also write the centrifugal term like a derivative of the potential term as has been done in the case of the term involving the acceleration of the frame, but it will obscure the covariance of the kinematic term, which could be easily constructed from the logic given in the previous section. Also, written in the form (\ref{kin}), we readily see that the acceleration of the frame mimics the effect of an uniform gravitational field. It is reminescent of the relativistic case where to achieve Weyl covariance we also promote ordinary derivatives to covariant derivatives which also conforms with the equivalence principle.
 
\subsection{The Pressure Term}
The pressure term in a non-inertial frame is just $-(\nabla_{i}p)/\rho$. We will see that the pressure term and even the viscous term requires no modification and by themselves transform covariantly under the full GCA. 

The pressure term is
\begin{equation}\label{pt}
-\frac{\nabla_{i} p}{\rho}.
\end{equation}
We make a natural assumption that the density transforms homogenously under GCA, so that 
\begin{equation}\label{density}
\rho(\mathbf{x},t) = (\frac{dt^{'}}{dt})^{a}\rho^{'}(\mathbf{x}^{'},t^{'}),
\end{equation}
where $a$, is an undetermined constant. Therefore the pressure term should remain
covariant if the pressure $p$ transforms in exactly the same manner as the density
$\rho$, so that
\begin{equation}\label{pressure}
p(\mathbf{x},t) = (\frac{dt^{'}}{dt})^{a}p^{'}(\mathbf{x}^{'},t^{'}).
\end{equation}
Finally one gets,
\begin{equation}
-\frac{\nabla_{i} p}{\rho} = -\frac{dt^{'}}{dt}R_{ij}^{-1}\frac{\nabla_{i}^{'} p^{'}}{\rho^{'}},
\end{equation}
as claimed.

\subsection{The Viscous Term}
The viscous term in non-inertial frame is:
\begin{equation}
-\frac{\nabla_{i}(\eta \sigma_{ij})}{\rho} =- \frac{\nabla_{i}\left(\eta \left(\nabla_{i}v_{j}+\nabla_{j}v_{i} -\frac{2}{3}\delta_{ij}(\mathbf{\nabla\cdot v})\right) \right)}{\rho}.
\end{equation}
We will see that this term is covariant by itself under the full GCA without any modification. We have already seen in (\ref{list}) that $\nabla_{i}$ and the traceless symmetric tensor $\sigma_{ij}$ both transform covariantly. We have already seen how the density field should transform in (\ref{density}). So clearly, the viscous term transforms like the kinematic term provided
\begin{equation}
\eta(\mathbf{x},t) = (\frac{dt^{'}}{dt})^{a-1}\eta^{'}(\mathbf{x}^{'},t^{'}).
\end{equation}
With the above rule for tranformation of the viscosity we get as desired.
\begin{equation}
-\frac{\nabla_{i}(\eta \sigma_{ij})}{\rho} = -\frac{dt^{'}}{dt}R_{jl}^{-1}\frac{\nabla_{k}^{'}(\eta^{'} \sigma_{kl}^{'})}{\rho^{'}}.
\end{equation}

\subsection{Summing all up}
The full covariant form of the Navier-Stokes equation is:
\begin{equation}
(\mathcal{D}v)^{(comb)}_{i} = -\frac{\nabla_{i}p}{\rho}- \frac{\nabla_{j}\left(\eta\left(\nabla_{i}v_{j}+\nabla_{j}v_{i} -\frac{2}{3}\delta_{ij}(\mathbf{\nabla\cdot v})\right)\right)}{\rho},
\end{equation}
or,
\begin{eqnarray}
\mathcal{D}(v_i + \Omega_{ij}(t)x_j - \frac{\mathbf{\nabla.v}}{d}x_i) + \Omega_{ij}(t)(v_j + \Omega_{jk}(t)x_k - \frac{\mathbf{\nabla.v}}{d}x_j) -\nabla_{i}(\mathbf{\mathcal{B}}(t).\mathbf{x}) \\\nonumber
=  -\frac{\nabla_{i}p}{\rho}- \frac{\nabla_{j}\left(\eta\left(\nabla_{i}v_{j}+\nabla_{j}v_{i} -\frac{2}{3}\delta_{ij}(\mathbf{\nabla.v})\right)\right)}{\rho}.
\end{eqnarray}
Besides, the density, pressure and viscosity transforms as follows,
\begin{eqnarray}\label{trans}
\rho(\mathbf{x},t) = (\frac{dt^{'}}{dt})^{a}\rho^{'}(\mathbf{x}^{'},t^{'}),\\\nonumber
p(\mathbf{x},t) = (\frac{dt^{'}}{dt})^{a}p^{'}(\mathbf{x}^{'},t^{'}),\\\nonumber
\eta(\mathbf{x},t) = (\frac{dt^{'}}{dt})^{a-1}\eta^{'}(\mathbf{x}^{'},t^{'}).
\end{eqnarray}

We will now investigate some interesting consequences of the above transformations. Let us first consider the case when all chemical potentials are zero as in
a gas of phonons in a metal. Then both the density and pressure are functions of
temperature, which must transform appropriately under GCA to reproduce (\ref{density})
and (\ref{pressure}). The speed of sound $c_{s}$ in the comoving frame (i.e in the local inertial
frame comoving with the local velocity v of the flow) is given by $c_{s}^{2}=dp/d\rho$. Since the pressure and density tranform identically under GCA, we find that $c_{s}$ is
invariant under GCA.
    
In a typical Galilean invariant theory this is not surprising, as for instance,
for monoatomic ideal gases, with molecular weight $m$, $c_{s} = \sqrt{(5kB T/3m)}$. The
temperature field being Galilean invariant, Galilean invariance of $c_{s}$ is automatic.
The problem is that a GCA invariant microscopic theory (as argued in \cite{0902.1385}) cannot
have any mass parameter. Here, the temperature T does transform non-trivially
under GCA, so $c_{s}$ must either be a fundamental constant like the speed of light or
be given in terms of the microscopic parameters of the theory. The situation is the
same in a relativistic conformal system where the speed of sound is $c/ \sqrt{3}$, where
$c$ is the speed of light.
    In a typical non-relativistic theory there is no fundamental speed. However,
there is a novel possibility, when the number of spatial dimensions is two. We
have seen that, in this case, the GCA admits a central charge, $\Theta$, which has the
dimension of $(1/speed)^2$ and also being a central charge, this is invariant under
GCA. So, in this case, we have a natural origin for a fundamental speed, which is
$1/ \sqrt{|\Theta|}$. In other dimesnions, $c_s$ must be given in terms of microscopic parameters,
for instance it can be the ratio of a microscopic length parameter and a microscopic
time parameter. We will have more to say about this possibility later.
    In any case, for a system without chemical potentials, cs must be a constant.
However, if we have chemical potentials too, cs need not be so and the analysis
above is insufficient to make any conclusion in this case.

\section{The influence of the continuity equation}
We will see here that the constant $a$, which governs the transformation of density
and pressure under the full GCA can be fixed uniquely by the continuity equation.
The continuity equation is
\begin{equation}\label{conteqn}
\mathcal{D}\rho + \rho (\mathbf{\nabla\cdot v}) = 0.
\end{equation}
Let us study how this equation transforms under the full GCA (say as represented in (\ref{comb}). We assume, as we did in the previous section that the density field transforms homogenously, so that
\begin{equation}
\rho(\mathbf{x},t) = (\frac{dt^{'}}{dt})^{a}\rho^{'}(\mathbf{x}^{'},t^{'}).
\end{equation}
With this assumption, we readily see that
\begin{equation}
\mathcal{D}\rho + \rho (\mathbf{\nabla.v}) = (\frac{dt^{'}}{dt})^{a+1}(\mathcal{D}^{'}\rho^{'} + \rho^{'} (\mathbf{\nabla}^{'}.\mathbf{v}^{'}))+\rho^{'}(\frac{dt^{'}}{dt})^{a-1}(\frac{d^{2}t^{'}}{dt^{2}})(a-d).
\end{equation}
So clearly we have covariance for the continuity equation only if $a=d$. So \textit{the continuity equation, if valid, predicts the transformation of the density under GCA}. 

We will see what consequences we now have for the Navier-Stokes' equation. If the pressure term has to be covariant under GCA and transform exactly like the kinematic term, we require that the pressure transforms in the same way as the density, so
\begin{equation}
p(\mathbf{x},t) = (\frac{dt^{'}}{dt})^{d}p^{'}(\mathbf{x}^{'},t^{'}).
\end{equation}
We immediately see that the pressure transforming in the same way as the density again makes the speed of sound $c_{s}$ a constant, when all chemical potentials vanish.

We now turn to the viscous term. Again we easily see that to achieve GCA covariance of the viscous term, we require that the viscosity transforms under GCA as below,
\begin{equation}\label{vis}
\eta(\mathbf{x},t) = (\frac{dt^{'}}{dt})^{d-1}\eta^{'}(\mathbf{x}^{'},t^{'}).
\end{equation} 

Finally, we note that if there is no particle number conservation the continuity
equation written in the form (\ref{conteqn}) should not hold. In this case the RHS must be
non-vanishing owing to say, particle absorption or emission. However, we will
still have the same conclusions as it will be natural to demand that the LHS of
this modified equation, which will be the same as before, must be covariant under
GCA on its own.

\section{GCA Covariance and the Viscosity}
The covariance of the Navier-Stokes equation and the continuity equation under
the full GCA requires that the viscosity should transform in a certain specified
manner as given by (\label{vis}). Now, the viscosity can transform only through its de-
pendence on the thermodynamic variables which are pressure and density. Here,
as before, we will assume the absence of chemical potentials. We note that $p/\rho$
does not transform under GCA as both the pressure and density transform exactly
the same way. So the only way, in which we can achieve the required transfor-
mation of the viscosity under the full GCA is that it depends on the pressure and
density in the following manner,
\begin{equation}
\eta = A(\frac{p}{\rho})^{x}p^{\frac{d-1}{d}},
\end{equation}
where $A$ is a \textit{dimensionful} microscopic parameter. The dimension of $A$ turns out to be:
\begin{equation}
[A] = M^{\frac{1}{d}}(\frac{L}{T})^{-\frac{d-2}{d}-2x}.
\end{equation}
In the equation above, $A$ is a (dimensionful) parameter and not a field, so it does not transform under GCA. It is a parameter because it is independent of the thermodynamic quantities like the pressure and density and of course it is independent of the velocity field as well. So, $A $ must be given by some
microscopic parameters and fundamental constants like the Planck's constant $h$.
However, as argued in \cite{0902.1385}, no microscopic theory which is GCA invariant, can
contain any mass parameter, so the mass dimension of A can come only through
the Planck's constant $h$. Without any loss of generality, we may also assume that
we have a length scale $l_f$ in the theory, which by definition is a parameter in the
theory and unlike the thermal wavelength this has no dependence on the temperature or any other thermodynamic variable by definition. Since generically
we do not have any fundamental speed like the speed of light in a non-relativistic
theory, we need an independent microscopic time scale $t_f$ also, which is again
by definition independent of thermodynamic variables, to soak the time dimension of $A$. We need an independent time scale in the microscopic theory, because unless there is a fundamental speed or a fundamental quantity with dimension of speed, we cannot form a time scale out of a length scale. Finally, without loss of generality, we can say that $A$ should take the form below
\begin{equation}
A \approx h^{\frac{1}{d}}l_{f}^{-1-2x}t_{f}^{\frac{d-1}{d}+2x}.
\end{equation}

It is clear from the above equation that we cannot make the dependence of $A$ on
the microscopic length scale $l_f$ and the microscopic time scale $t_f$ vanish simultaneously. Therefore, we conclude that we can explain the required transformation
of the viscosity under the full GCA only if we have a microscopic length scale
or a microscopic time scale or both in our theory. We also note that even when
$d = 2$, in which case the central $\Theta$ allows to define a "fundamental speed," given
by $\sqrt{1/ |\Theta|}$, it is impossible to soak the dimension of $A$ with the Planck's constant
and $\Theta$ alone. So it is impossible to do without introducing a microscopic length
scale or microscopic time scale or both.

The conclusion, therefore, is that in a GCA invariant theory, either the viscosity is zero or it contains a microscopic length parameter or a microscopic time
parameter or both. This is indeed contrary to the case of a relativistic conformal
field theory where we cannot have any intrinsic length parameter or time parameter and any quantity can have a dimension only through the Planck's constant and
the speed of light. At this moment, we do not know any GCA invariant microscopic theory so we can be open to the possibility that such theories can contain
intrinsic length or time parameters or both. If this is not possible, then the viscosity should vanish. Of course, as in the case with our analysis of $c_{s}$ , our conclusions
may change if we introduce chemical potentials.

One may however, ponder if it is possible that GCA could be a symmetry of the theory only in the presence of non-zero chemical potentials so that the above considerations for the case of vanishing chemical potentials can be avoided. In our opinion, this point of view is rather unnatural, because the symmetry of a theory is usually a fundamental property of the theory and though its manifestation might be modified, it can neither appear or disappear at specific values of thermodynamic intensive variables like temperature or chemical potentials. An easy example which supports this point is the usual relativistic conformal symmetry of $\mathcal{N}=4$ SYM theory, in which case in presence of a finite temperature we still have conformal symmetry, however the thermodynamic variables also transform under conformal transformations. In the Discussion section, we will point out possible significances of the analysis done here in the case of vanishing chemical potentials for AdS/CFT realization of GCA.

\section{Possible GCA covariant corrections to the Navier-Stokes Equation}

The Navier-Stokes equation, being a phenomenological equation, is succeptible to higher derivative corrections, which could be, in principle, calculated from kinetic theory. We will see that GCA is powerful in constraining these corrections, quite like in the case of hydrodynamics covariant under the relativistic conformal group. So, this will give us further evidence, that GCA indeed is a credible physical symmetry, that is a symmetry which can constrain phenomenological laws (in absence of known GCA invariant microscopic theories). \footnote{The author would like to thank Rajesh Gopakumar for pointing out this significance of the constraints imposed by GCA on the correction to the Navier-Stokes' equation.}

Usually, for instance, if calculated from the kinetic theory of gases, the corrections to the Navier-Stokes involve corrections to the dissipative part of the stress tensor $\tau_{ij}$, which at the first-order in derivatives is just $\eta \sigma_{ij}$. The next-order corrections to the Navier-Stokes equation are contained in the two derivative corrections, $\tau^{(2)}_{ij}$, to the dissipative stress tensor, so that $\tau_{ij} = \eta \sigma_{ij}+ \tau^{(2)}_{ij}$ and the corrected Navier-Stokes' equation in the inertial frame, now takes the form,
\begin{equation}
\mathcal{D}v_{i} = -\frac{\nabla_{i} p}{\rho} - \nabla_{i} (\tau_{ij}) = -\frac{\nabla_{i} p}{\rho} - \nabla _{i} (\eta \sigma_{ij}+ \tau^{(2)}_{ij}).
\end{equation}
Now, we would demand that like $\sigma_{ij}$, $\tau^{(2)}_{ij}$ contains spatial derivatives only as is indeed that case if these corrections are calculated from kinetic theory. Also, we will assume, that these corrections involve derivatives of the velocity only. 

Let us first look at terms in $\tau^{(2)}_{ij}$ which have the structure of $(\nabla u)^{2}$. For that, we need to find if there is any other tensor with structure $(\nabla u)$ which transforms like $\sigma_{ij}$. One can easily see that there is only one more such tensor, which we denote as $\omega_{ij}$ and is defined as below
\begin{equation}
\omega_{ij} = \frac{1}{2}(\nabla_{i} u_{j} - \nabla_{j} u_{i} - 2\Omega_{ij}(t))
\end{equation}
Once again by invoking the trick of comparing one inertial frame with two non-inertial frames and then comparing the two non-inertial frames with each other one can readily prove that $\omega_{ij}$ transforms under full GCA like $\sigma_{ij}$. Therefore $\tau^{(2)}_{ij}$ involve the following combinations $\lambda_{1} \sigma_{ik}\sigma_{kj} + \lambda_{2} (\sigma_{ik}\omega_{kj}+ \omega_{ik}\sigma_{kj}) + \lambda_{3} \omega_{ik}\omega_{kj}$, where the three $\lambda$'s are arbitrary transport coefficients like the shear viscosity $\eta$. For the covariance of the corrected Navier-Stokes we now require them to transform as below,
\begin{equation}\label{lambda}
\lambda_{i}(\mathbf{x},t) = (\frac{dt^{'}}{dt})^{a-2}\lambda_{i}^{'}(\mathbf{x}^{'},t^{'})
\end{equation}
where $i = 1,2,3$ and $a$ is defined through the transformation of the density as given in (\ref{density}). We can proceed to find the dependence of the $\lambda$'s on the thermodynamic variables exactly as we have done for the shear viscosity $\eta$, however we will not repeat it here.

Now let us look for possible corrections to $\tau^{(2)}_{ij}$ which contains the structure $(\nabla^{2}u)$. Now since $\mathbf{v.\nabla}$ does not transform covariantly, we cannot try combinations like $\mathbf{(v.\nabla)}\sigma_{ij}$. Moreover, though the Laplacian, $\Box$, transforms covariantly, we cannot use it on any polynomial of the velocity like $u_{i}u_{j}$, as it is not covariant. It is not, thus hard to see, that there is only one possible covariant term which contains a $(\nabla^{2}u)$ term and it is $\nabla_{k}(\sigma_{ij}\mathcal{V}^{(\mathbf{b}=0)}_{k})$, where $\mathcal{V}^{(\mathbf{b}=0)}_{k}$ is as defined in (\ref{covvb=0}). We can still get a covariant term, though $\mathcal{V}^{(\mathbf{b}=0)}_{k}$ is covariant only in absence of boosts, because the full covariant velocity field will differ from this by a purely time-dependent quantity, so it doesn't make any difference when we apply the spatial derivative. We note that, in an inertial frame, however, this new term is just $\mathbf{(v.\nabla)}\sigma_{ij}$. We will denote the coefficient corresponding to this term as $\lambda_{0}$.

Therefore, the most general form of $\tau^{(2)}_{ij}$ is:
\begin{equation}
\tau^{(2)}_{ij} = \lambda_{0}\nabla_{k}(\sigma_{ij}\mathcal{V}^{(\mathbf{b}=0)}_{k}) + \lambda_{1} \epsilon_{ik}\epsilon_{kj} + \lambda_{2} (\epsilon_{ik}\omega_{kj}+ \omega_{ik}\epsilon_{kj}) + \lambda_{3} \omega_{ik}\omega_{kj},
\end{equation}
with all $\lambda$'s having appropriate dependence on thermodynamic variables so that it transforms as in (\ref{lambda}). 

Similarly, we can proceed to constrain higher order corrections of the Navier-Stokes' equation containing more than three derivatives. We observe that our four possible GCA covariant corrections, have analogues in the relativistic conformal case, as all the four possible corrections in flat space-time \cite{Baier:2007ix}, reduce in the non-relativistic limit to our four terms in an inertial frame when the flow is incompressible. This is intriguing because the covariant forms in the two cases are very different in content. It will be interesting to see if this correspondence also exist at higher orders. There can be another term in our case involving the curvature of the spatial metric as in the relativistic case (the relativistic term involves contractions of the Reimann tensor), but since we have throughout restricted ourselves to the flat spatial metric, this possibility lies outside the scope of our present investigation.

\section{Discussion}
We have shown that the macroscopic Navier-Stokes equation for incompressible
flows has covariance under full GCA. So we can conclude that GCA can be realized as a symmetry of a phenomenological law like the Navier-Stokes equation only if we covariantize the usual form of the laws which holds in inertial
frames, however not any arbitrary law with mere Galilean covariance can be covariantized. In the case of the Navier-Stokes equation we have needed that the
flow is incompressible. We have also seen that the higher derivative corrections to
the Navier-Stokes equation can be constrained by requiring GCA covariance.

Our analysis also leads us to conclude that when all chemical potentials vanish,
$c_s$, which denotes the speed of sound in a comoving frame, is a constant. Further,
we have seen that in the absence of chemical potentials, the viscosity should either
vanish or in the microscopic theory we must have a length scale or a time scale or
both.

We would now like to discuss the possible implications of the above analysis for AdS/CFT realization of GCA. The presence of both length and time scales in the GCA invariant microscopic theory firstly tallies with the fact that we need to introduce objects like absolute angular velocity and absolute acceleration of the non-inertial frame which brings in dimensions of both length and time into play. This is in contrast with the case of covariantizing under relativistic conformal group where we need not bring in any additional dimensionful parameter. This observation possibly indicates that we need to first deform the action of the relativistic parent theory like $\mathcal{N}=4$ SYM by \textit{non-marginal} operators such that a deformed $SO(d,2)$ relativistic conformal group is the symmetry of the theory and then take the contraction which takes $SO(d,2)$ relativistic conformal group to GCA so that we get a sensible dynamical limit \footnote{A related example could be the omega-deformation \cite{Nekrasov} of $\mathcal{N}=2$ SYM theories under which the deformed theory retains the BRST supersymmetry though this supersymmetry itself gets deformed by combining with other supersymmetries.}. The deformation parameters of the symmetry being dimensionful, should bring in the required microscopic length scales and time scales in the final GCA invariant theory obtained via the contraction. Further, the deformation parameters will also transform non-trivially under GCA so that the covariantizing will bring in new structures. In fact, if we take the contraction without deformation for the classical $N=4$ SYM theory, one may readily check that we get a non-dynamical equations of motion for all the fields \footnote{The author thanks Rajesh Gopakumar for valuable discussions regarding these points.}. This supports our point of view.  In the future, we would like to find out the appropriate operators which could give rise to the deformations such that the contraction produces a sensible dynamical theory.

Finally, we mention, that it would be an interesting challenge to construct
gravitational duals for GCA covariant hydrodynamic flows. Aside from finding
the dynamics of gravity in the bulk, we see now, we also need to find a suitable
bulk interpretation of the absolute angular velocity and the absolute acceleration
of the boundary coordinate system, as they are surely needed in the covariant formulation of the hydrodynamics of the boundary theory. Some earlier work in \cite{Plyuschay} could be useful in this direction.

\textbf{Acknowledgments:} The author would like to thank Rajesh Gopakumar for valuable discussions and also for meticulously checking the manuscript. He would like to thank Arjun Bagchi for his several suggestions on improving the manuscript and Yogesh Srivastav for useful discussions.                                     He also thanks the hospitality of
IMSc, Chennai and CHEP, IISc, Bangalore where some parts of this work have
been done. Finally, the support of the people of India for research in basic sciences is reverently acknowledged.
\section*{Appendix A: A Simple Mathematical Interpretation of the GCA}
Mathematically, the infinite dimensional GCA can be motivated as follows: \textit{Consider two particles with velocities $\mathbf{v_1}$ and $\mathbf{v_2}$ respectively at the same point in space $\mathbf{x}$ and at the same time $t$. Then the infinite dimensional GCA is the largest possible group of space-time transformations under which the relative velocity $(\mathbf{v_1 - v_2})$ transforms covariantly (as a vector under rotation) while its norm remains invariant.}. We will now prove this statement.

Let us consider an arbitrary space-time transformation from $(\mathbf{x}, t)$ to $(\mathbf{x^{'}}, t^{'})$. Let us denote
\begin{equation}
M_{ij}=\frac{\partial x^{'}_{i}}{\partial x_{j}}, N_{i} = \frac{\partial x^{'}_{i}}{\partial t}, P_{i} =\frac{\partial t^{'}}{\partial x_{j}}, Q =\frac{\partial t^{'}}{\partial t}.
\end{equation}
Then the following holds,
\begin{eqnarray}
dx^{'}_{i} = M_{ij}dx_{j} + N_{i}dt,\\\nonumber
dt^{'} = P_{i}dx_{i} + Qdt.
\end{eqnarray}
So, we have
\begin{equation}
v^{'}_{i} = \frac{M_{ij}v_{j} + N_{i}}{P_{k}v^{k} + Q}.
\end{equation}
The relative velocity of two particles at the same point in space at a given time transforms as below,
\begin{equation}
v^{'}_{(1)i} - v^{'}_{(2)i} = \frac{(M_{ij}v_{(1)j}P_{k}v_{(2)k}-M_{ij}v_{(2)j}P_{k}v_{(1)k}) + Q(M_{ij}v_{(1)j}-M_{ij}v_{(2)j})+ N_{i}(P_{k}v_{(2)k}-P_{k}v_{(1)k})}{(P_{l}v_{(1)l} + Q)(P_{m}v_{(2)m} + Q)}.
\end{equation}
For this transformation to be covariant, we require $P_{k} = 0$, in which case
\begin{equation}
v^{'}_{(1)i} - v^{'}_{(2)i} = \frac{M_{ij}v_{(1)j}-M_{ij}v_{(2)j}}{Q}.
\end{equation} 
If we also require the norm to remain the same, we should have,
\begin{equation}
\frac{M_{ij}}{Q} = R_{ij},
\end{equation}
where, $R_{ij}$ is a rotation matrix. Now, $P_{i} = (\partial t^{'}/\partial x_{i}) =0$ implies that 
\begin{equation}\label{time}
t^{'} = f(t), Q= \frac{df(t)}{dt}.
\end{equation}
Then we have 
\begin{equation}
M_{ij} = \frac{\partial x^{'}_{i}}{\partial x_{j}} = Q R_{ij}(\mathbf{x}, t) = \frac{df(t)}{dt} R_{ij}(\mathbf{x}, t).
\end{equation}
The integrability condition requires that
\begin{equation}
\frac{\partial M_{ij}}{\partial x_{k}} = \frac{\partial M_{ik}}{\partial x_{j}},
\end{equation}
which in turn implies that
\begin{equation}
\frac{\partial R_{ij}(\mathbf{x},t)}{\partial x_{k}} = \frac{\partial R_{ik}(\mathbf{x},t)}{\partial x_{j}}.
\end{equation}
The above condition at a fixed value of $i$, the implies that the curl of a vector vanishing so that we must have
\begin{equation}\label{par}
R_{ij}(\mathbf{x},t) = \frac{\partial V_{i}(\mathbf{x},t)}{\partial x_{j}}.
\end{equation}
A rotation matrix satisfies the property that $R^{-1}_{ij}=R_{ji}$, so we should have
\begin{equation}
\frac{\partial V_{i}}{\partial x_{j}}\frac{\partial V_{k}}{\partial x_{j}}=\delta_{ik}.
\end{equation}
The solution to the above system of equations is\\
$V_{i} = \overline{R}_{ij}(t)x_{j} +$ a function of time,\\
so, we have $R_{ij} = \overline{R}_{ij}(t)$. To sum up, $(\partial x^{'}_{i}/\partial x_{j}) = QM_{ij} = (df(t)/dt)R_{ij}(t)$, therefore
\begin{equation}\label{space}
x^{'}_{i} = \frac{df(t)}{dt}R_{ij}(t) x_{j} + b_{i}(t).
\end{equation}
The above together with (\ref{time}) belongs to our group of spacetime transformations denoted by GCA. 

It is also easy to check that any transformation belonging to the GCA makes the relative velocity of two particles at a given point in space at a given time transform covariantly while preserving its norm.  So we have proved, that the largest group of spacetime transformations under which the relative velocity of two particles at the same point in space at a given time transforms covariantly while its norm is preserved, is the GCA. This mathematical result can have physical applications in constructing local interactions of particles in a GCA-invariant microscopic theory.

\section*{Appendix B: $G = MLR$}
Here, we will prove that any arbitrary element ($G$) of GCA, can be written uniquely as a succession of a time dependent rotation ($R$), a spatially correlated time reparametrisation ($L$) and a time dependent boost ($M$). 

Let us denote the space-time coordinates $(\mathbf{x},t)$ together as $X$. Let $G$ be an arbitrary element of the GCA and let two coordinates $X$ and $X^{'}$ be related so that $X^{'} = G.X$, i.e. $X^{'}$ is the result of action of $G$ on $X$. 

However, we now note that there is a \textit{unique} time-dependent boost $M$ such that $M.X$ and $X^{'}$  will \textit{will share the same origin of spatial coordinates at all times}. Let us denote $M^{-1}.X^{'}$ as $X^{''}$. So, by construction $X^{''}$ and $X$ share the same origin of spatial coordinates \textit{at all times}. 

Now, if two space-time coordinates share the same origin of spatial coordinates at all times, it is also easy to see, that there is a \textit{unique} spatially correlated time reparametrisation $L$ which relate their times. Therefore, there is a \textit{unique} $L$ such that $ X^{'''} = L^{-1}.X^{''}$ and $X$ share the same time.

By construction, we see that $X^{'''}$  and $X$ share the same time and the same origin of spatial coordinates. Therefore, they must be related by a \textit{unique} time-dependent rotaion $R$, so that $X = R^{-1}.X^{'''}$.

Summing all up, $X = R^{-1}.X^{'''} = R^{-1}L^{-1}X^{''} = R^{-1}L^{-1}M^{-1}X^{'}$. But we assumed $X = GX^{'}$, so $G = MLR$, with $M$, $L$ and $R$ being unique because they were unique in each stage of our argument above. So, we have proved that any arbitrary element ($G$) of GCA, can be written as a succession of a time dependent rotation ($R$), a spatially correlated time reparametrisation ($L$) and a time dependent boost ($M$).

\providecommand{\href}[2]{#2}\begingroup\raggedright

\end{document}